\documentstyle[11pt,aasms4]{article}
\begin{document}
   \title{Short term pulse frequency fluctuations of OAO 1657-415 from
RXTE observations  }

   \author{Altan Baykal 
          }

\affil {
 Physics Department, Middle East Technical University,
Ankara 06531, Turkey}

   \begin{abstract}

We present new X-ray observations of the high mass X-ray binary 
(HMXRB)
pulsar OAO 1657-415, obtained 
during one orbital period (10.44 days) 
with the Rossi X-Ray Timing Explorer (RXTE).
Using the binary orbital parameters, obtained from  
Burst and Transient Source Experiment (BATSE) observations, 
we resolve the fluctuations in the pulse frequency at time scales 
on the order of one day for the first time.
Recent BATSE results showed that OAO 1657-415
has spin-up/down trends in its pulse frequency time series,
without any correlation with the X-luminosity at
energies $>$20 keV (Baykal 1997).
In the present RXTE observations 
the source is found to be in an extended phase of spin-down. 
We also find a gradual increase in the X-ray luminosity which 
is correlated with a marginal spin-up episode. 
The marginal correlation between 
the gradual spin-up (or decrease in spin-down rate)
and increase in X-ray luminosity suggests that  
the OAO 1657-415 is observed during  
a stable accretion episode where the 
prograde accretion disk is formed. 


\end{abstract} 
\section{Introduction}

The high mass X-ray binary source HMXRB source
 OAO 1657-415 (OAO 1653-40) was first detected by the Copernicus
satellite (Polidan et al. 1978) in the 4-9 keV range. 
 The HEAO-1 observations also
showed 38.22 sec pulsations in the 1-40 keV and
40-80 keV bands (White $\& $ Pravdo 1979, Byrne et al. 1981).
Observations
with Ginga and GRANAT
(Kamata et al. 1990, Gilfanov et al. 1991,
Mereghetti et al. 1991, Sunyaev et al. 1991)
have found pulse period changes. 
BATSE observations of this source with the
Compton Gamma Ray Observatory (CGRO) 
showed that OAO 1657-415 is in an $10.44^{d}$ binary
orbit with an X-ray eclipse by the a stellar companion
(Chakrabarty et al. 1993). 
The observed orbital parameters
imply that the companion is a supergiant of
spectral class B0-B6.
The correlations between X-ray flux and pulse frequency derivatives   
($\dot \nu $) fluctuations were investigated  
by using the previously published pulse frequencies 
and BATSE measurements (Baykal 1997). 
These correlations     
suggested that the formation 
of episodic accretion disks in the case of a stellar wind 
is the possible accretion mechanism. 

In this paper, we present the short term pulse frequency fluctuations 
and X-ray fluxes 
of OAO 1657-415 
in the light of recent RXTE observations. 
We have employed background 
subtraction by using the background models for the RXTE/PCA instrument 
and galactic ridge emission in the 2-50 keV range. Our X-ray flux and pulse 
frequency measurements find an increase in the X-ray flux which 
is correlated with the decrease in the spin-down rate (or marginal 
spin-up trend).

\section{Observation and Data Analysis}

OAO 1657-415 was observed between 1997 August 20-27 
within the guest observer program of RXTE
with proposal observation  
ID 20113. 
RXTE pointings of the source are  
separated from each other by approximately six hours 
with a total observation span of 75 ksec.
The results  
presented here are based on data collected with the Proportional Counter
Array (PCA, Jahoda et al., 1996) and the High Energy X-ray Timing Experiment
(HEXTE, Rothschild, et al., 1998). The PCA instrument consists of an array of
5 proportional counters operating in the 2--60 keV energy range, with a
total effective area of approximately 7000 cm$^2$ and a field of view
$\sim 1^{\circ}$ FWHM. The HEXTE instrument consists of two independent clusters of 
detectors, each cluster containing four NaI(T1)/CsI(Na) phoswich scintillation 
counters sharing a common $\sim 1^{\circ}$ FWHM field of view. 
The field of view of each cluster is switched 
on and off source to provide background measurements. The net open 
area of the seven detectors is 1400 cm$^2$ and each detector covers 
the energy range 15--250 keV.

\subsection{X-ray Light-Curves and Spectra}

Background light-curves and  
pulse height amplitudes are generated by using the 
background estimator models based upon the rate of very large events (VLE),
spacecraft activation and cosmic X-ray emission with the standard PCA
analysis tools. The background light-curves are subtracted 
from the source light-curve obtained from the Good Xenon event data 
(Fig. 1). 
 Since the source is close to the galactic center, 
galactic ridge data were 
extracted from archival RXTE observations. Observations 
pointed at directions a few degrees away from the source are collected. 
The background spectra for the galactic ridge data were generated using 
PCA analysis tools.  
After the instrument background are removed from the galactic ridge data, 
the residual spectra are used as a background for the X-ray spectra  
of OAO 1657-415. The source spectrum was calculated using 
the same PCA background estimator models. 
(It should be noted that   
$\sim 3$ $\%$ systematic errors are used
 in spectral fitting of PCA data). 
For the HEXTE data the background subtraction is straightforward since 
the HEXTE detectors are rocking in 16 sec intervals. Standard ftools software 
for RXTE is used for the data reduction and for the dead time correction.
We have used power-laws with
 high energy cut-off models together with a 6.65 keV 
gaussian emission line which was found in 
previous GINGA observations
(Kamata et al., 1990). We have found no evidence for any deviation from 
a power-law model with high energy cut off which might be attributable 
to a cyclotron feature.   
Table 1 presents the spectral parameters of X-ray spectra. 
Figure 2 presents the joint X-ray spectra of RXTE/PCA and HEXTE detectors.
The power-law index, and the cut-off and e-folding energies are
significantly different from those measured by Ginga, the powerlaw steeper,
but extending to higher energy. The broader energy band of the PCA and
HEXTE combination better constrain the higher energy spectra, but the
spectrum may vary intrinsically.

\subsection{Pulse Timing of OAO 1657-415}

The background subtracted light-curves are corrected with respect to the 
barycenter of the solar system. Using the binary orbital parameters of 
OAO 1657-415 from BATSE observations (Bildsten et al., 1997), 
the light curves are 
also corrected for binary motion of OAO 1657-415 (see Table 2).
A long power spectrum 
was used to estimate the average pulse frequency.
This 
pulse frequency is consistent
 with BATSE pulse frequency records at the same time 
(obtained through HEASARC http:cossc.gsfc.nasa.gov).
 In order to resolve the pulse arrival times and 
pulse frequencies at shorter 
time scales 
29 pulse arrival times were generated (one pulse arrival time for each RXTE 
orbit). 
Pulse arrival times are found 
by folding the light-curve data into one average pulse 
for each RXTE orbit, folding all light-curves into one master pulse, 
and cross-correlating the master pulse with each of the 29 average 
pulses. In the pulse timing anaysis, we have used the method of 
harmonic representation of pulse profiles, as proposed by
Deeter $\&$ Boynton (1985). In this method, pulse profiles 
are expressed in terms of harmonic series and cross-correlated
with the master pulse profile. The maximum value 
of the cross-correlation is analytically well defined and does not depend  
on the phase binning of the pulses.
The master pulse with 40 phase
bins was represented by their harmonics (Deeter $\&$ Boynton 1985)
and cross-correlated with harmonic representations of pulse profiles
from segments of the data.

The pulse profiles for OAO 1657-415 are variable. This affects the pulse 
timing.
In order to estimate the errors in the arrival times,  
the light-curve of each RXTE orbit 
is divided into approximately 10-15 equal 
subsets and new arrival times are estimated.
The average variance in the arrival times are computed and treated 
as errors of arrival times. 
The pulse arrival 
times are represented in Fig. 3. The residual   
pulse arrival times may arise from the 
change of the pulse frequency during the observation (or
intrinsic pulse frequency derivative) and from the errors of obital 
parameters (Deeter et al., 1981),
\begin{equation}
\delta \phi = \phi_{o} + \delta \nu (t-t_{o})
+  \frac{1}{2} \dot \nu (t-t_{o})^{2}
+ \nu \delta (\frac{asin i}{c}) sin l_{n}
- \nu \frac{2\pi \delta T_{\pi/2}}{P_{orbit}} \frac{asin i}{c} cos l_{n}
+ \nu \frac{2\pi}{P_{orbit}^{2}} \frac{asin i}{c} \delta P_{orbit}
  (t_{n}-T_{\pi/2}) cos l_{n}
\end{equation}
where $\delta \phi $ is the pulse phase offset deduced from the pulse
timing analysis,  $t_{o}$ is the mid-time of the observation,  $\phi_{o}$ is
the phase offset at t$_{o}$,  $\delta \nu$ is the deviation from the mean
pulse frequency (or additive correction to the pulse frequency),  and  $\dot
\nu $ is the pulse frequency derivative of the source, 
$\frac{asin i}{c}$ is the light travel time for 
projected semimajor axis, $T_{\pi/2}$ is the epoch when the mean orbital
longitude is equal to 90 degrees, $P_{orbit}$ is the orbital period, 
$\delta$ denotes the errors of these parameters, 
$l_{n}=2\pi (t_{n}-T_{\pi/2})/P_{orbit} +\pi/2$ is the mean orbital 
longitude at $t_{n}$. The above
 expression is fitted to the  pulse arrival times data. 
 Table 2 presents
the timing solution of OAO 1657-415 from our RXTE observations.  The pulse
frequency derivative obtained from the quadratic trend of the pulse timing
analysis is $\dot \nu _{RXTE} = -(3.27 \pm 0.09)\times 10^{-12} $ Hz
s$^{-1}$.
The average pulse frequency derivative we
deduce from the pulse frequency history of BATSE archival data
is $-(3.1 \pm 0.2)\times 10^{-12} $ Hz s$^{-1}$,
consistent 
 at the 1$\sigma$ level.

 The pulse frequency records of OAO 1657-415 
have shown that 
the source has stochastic spin-up/down trends (Baykal 1997, Bildsten  et al. 
1997) 
at timescales longer than weeks.
In intervals between stochastic changes,
OAO 1657-415 shows
secular spin-up or down trends with lower values of noise strength
(Baykal 1997).
 Our observations detect the source spinning down,  
on average. There are local deviations 
from quadratic trends which can be interpreted as very short term 
fluctuations (see Fig. 3).
 The torque noise analysis of OAO 1657-415 from the BATSE 
observations (Baykal 1997) showed that pulse frequency 
fluctuations at shorter than $\sim 8$ days are almost not detectable 
due to the measuremental noise, however RXTE observations are yielding 
significant fluctuations around the average quadratic 
trend (or spin-down rate), as shown in Fig. 3.
In RXTE observations we are able to construct pulses  
for time intervals as short as a few hundred seconds.
 This yields better timing at 
shorter time scales. Therefore the fluctuations in arrival times 
less than $\sim 8$ days are 
resolved.

\subsection{Torque and X-ray luminosity changes of OAO 1657-415}

The X-ray flux and pulse frequency derivative 
correlations of OAO 1657-415  
were investigated by using BATSE archival data base 
(Baykal 1997). 
BATSE pulse flux at 20-60 keV and pulse frequency series have shown 
no correlation between X-ray flux and pulse frequency derivatives.  
A strong correlation between specific angular momentum (l) 
and pulse frequency derivatives was found instead.
 These correlations implied that 
the specific angular momentum is directional, sometimes 
positive and sometimes negative ($\pm $ l), and that sometimes the flow is 
radial. These results suggested the formation of temporary 
accretion disks in the case 
of stellar wind accretion and the short term disk reversals are quite possible.
In the analysis of BATSE data, the shortest time scales for resolving 
the significant pulse frequency 
fluctuations were of the order of 8 days. 

In the present work, in order to resolve 
pulse frequencies at shorter time scales,   
we used high counting statistics of RXTE/PCA detectors 
and obtained pulse arrival times at approximately six hours intervals.
For each 4 or 5 pulse arrival times, we fitted a straight line 
segment and using the 
slope of this line we estimated 
a correction to the average pulse frequency  
($ \delta \phi = \phi_{o} + \delta \nu (t-t_{o}) $).  
In this way, we obtained pulse frequency 
records for each day roughly.
In Fig. 4 we present the pulse frequency records estimated in this work.
We accumulated the spectral data at 2-50 keV, corresponding to 
observations of pulse arrival times. In each set of observations we 
fitted a power-law with high energy cut-off with a gaussian 
emission line, then we estimate 
unabsorbed flux (or intrinsic pulsar flux). Fig. 5 presents the              
pulsar flux history during the RXTE observations. During the observation, 
bolometric X-ray flux increased roughly by 70$\%$.   
The spin-down rate is decreased and a marginal spin-up trend is seen. 
This is the first evidence in OAO 1657-415 to show 
positive marginal correlation between the X-ray flux and pulse frequency 
changes.

\section{Conclusion} 

OAO 1657-415, has shown strong spin-up/down torques in its 
time history which can not be explained by 
wind accretion (Baykal 1997).
The formation of temporary accretion 
disks was therefore considered.           
If accretion onto the neutron star is 
from a Keplerian disk (Ghosh \& Lamb 1979), the torque on the neutron star
is given by
\begin{equation}
I\dot \nu  = n(w_{s})  \dot M~l_{K},
\end{equation}
where $l_{K} = (GMr_{o})^{1/2}$ is the specific angular momentum 
added by a 
Keplerian disk to the neutron star 
at the inner disk edge $r_{o} \approx
0.5 r_{A}$ where $r_{A} \ = (2GM)^{-1/7} \mu ^{4/7} \dot M^{-2/7}$ is the
Alfven radius, $\mu$ is the neutron star magnetic moment, $n(w_{s}) \approx
1.4 (1-w_{s}/w_{c})/(1-w_{s})$ is a dimensionless function that measures
the variation of the accretion torque as estimated by the fastness
parameter $w_{s}=\nu  /\nu _{K}(r_{o}) = 2 \pi P^{-1} G^{-1/2}
M^{-5/7} \mu ^{6/7} \dot M^{-3/7}$. Here $w_{c}$ is the critical fastness
parameter where the accretion torque is expected to vanish at $w_{c} \sim
0.35-0.85$ depending on the structure of the disk. 
According to this model 
 the torque will cause a spin-up if the neutron star is rotating
slowly ($w_{s}~<~w_{c}$) in the same sense as the circulation in the disk,
or spin-down, if it is rotating in the opposite sense.
Even if the neutron star is rotating in the same sense as the disk flow,
the torque will spin-down the neutron star if it is rotating too rapidly
($w_{s}~>>~w_{c}$).  In such a model one should see positive correlation
between pulse frequency derivative ($\dot \nu $) and 
moderate mass accretion rate
($\dot M$) if the disk is rotating in the same sense as the neutron
star (Baykal 1997). 

Recent observations of accreting neutron stars have shown stochastic
spin-up/down trends on time scales from days to a few years (Bildsten et
al. 1997). 
Some of the sources switch from spin-up to spin-down states without
showing great changes in their mass accretion rates (Bildsten et
al. 1997).
These unusual behaviors led Nelson et al.
(1997) to the possibility of retrograde circulation of accretion disks. GX
4+1 shows correlation between the X-ray flux and the spin-down rate
(Chakrabarty et al. 1997) which may suggest a retrograde accretion disk.
In our RXTE observations, OAO 1657-415 has shown marginal correlation 
with accretion rate and pulse frequency change. This 
positive correlation strongly suggesting that the disk formed 
in the spin-down episode is in prograde direction. 
From the BATSE observations, it was concluded that 
the pulse frequency derivatives and X-ray flux 
were not well correlated and the 
8 days was the minimum for the 
correlation time scale (Baykal 1997).
This RXTE observation 
implies that the time-scale of correlation is short, 
only a few days. 
To see the exact nature of correlations between X-ray flux 
and pulse frequency derivatives, an even more   
extensive broad band X-ray observation should be carried out.

\begin{acknowledgements}
A.B. thanks 
Ali Alpar and Mark Finger for critical reading of the manuscript 
and  
Jean .H. Swank, Tod .E. Strohmayer, 
Mike Stark, 
for stimulating discussions
and USRA for supporting a visit to the
GSFC. 
\end{acknowledgements}

\newpage
 
\begin{table}
\caption{Spectral Fit Parameters for RXTE observations$^{a}$}
\label{Pri}
\[
\begin{tabular}{c|c}   \hline
N$_{H}$ (10$^{21}$cm$^{-2}$) & 12.94 $\pm$ 0.38 \\
Fe line center energy (keV) & 6.65 $\pm$ 0.03 \\
Fe line width (keV) & 0.40 $\pm$ 0.06 \\
Fe line intensity photons cm$^{-2}$ sec $^{-1}$ & (2.77 $\pm$ 0.31)$\times$
 $10^{-3}$ \\
Photon Index & 1.07 $\pm$ 0.03 \\
Cutoff Energy (keV) & 12.82 $\pm$ 0.29 \\
Folding Energy (keV) & 29.37 $\pm$ 0.98 \\
Power-law Norm. cm$^{-2}$ at 1keV & (5.96 $\pm$
0.37)$\times$ $10^{-2}$  \\
Reduced $\chi^2$ & 1.28 (d.o.f. = 178) \\ \hline
\end{tabular}
\]
$^{a}$ Uncertainties in the spectral fit
parameters denote single parameter 1 $\sigma $ errors.
\end{table}

\begin{table}
\caption{Timing Solution of OAO 1657-415 for RXTE Observations$^{a}:$
   }
\label{Pri}
\[
\begin{tabular}{c|c}  \hline
Orbital Epoch (MJD) & 48515.99(5)$^{b}$ \\
P$_{orb}$ (days)    & 10.44809(30)$^{b}$ \\
a$_{x}$ sin i (lt-sec) & 106.0(5)$^{b}$  \\
e                       & 0.104(5)$^{b}$  \\
w   & 93(5)$^{b}$ \\ 
Epoch(MJD)  & 50683.95400(2)      \\
Pulse Frequency (Hz) & 0.026775618(4) \\

Pulse Freq. Derivative (Hz s$^{-1}$)  &
-3.27(9)$\times 10^{-12}$ \\ \hline
\end{tabular}
\]
$^{a}$ Confidence intervals are quoted at the 1 $\sigma $ level. \\
$^{b}$ Orbital parameters are taken from Bildsten et al., (1997).
P$_{orb}$=orbital period, a$_{x}$ sin i=projected semimajor axis,
e=eccentricity, w=longitude of periastron.\\
\end{table}

{\bf Figure Caption}\\

Figure 1. The total(5 PCU) RXTE/PCA background subracted
X-ray light-curve in 2-50 keV energy range.

Figure 2. RXTE/PCA-HEXTE spectra of OAO 1657-415  
(Note that HEXTE spectra is the summed spectra of Cluster 1 and Cluster 2).
Below panel is the residuals of the fit in terms of 
$\chi ^{2}$ values.

Figure 3. Phase offsets in pulse arrival times. Solid line denotes the best fit 
of arrival times.

Figure 4. Pulse frequency measurements of OAO 1657-415,
                 from RXTE/PCA observations.

Figure 5.  2-50 keV X-ray flux measurements of OAO 1657-415,
                 from RXTE PCA observations.

\newpage
\clearpage
\begin{figure}
\plotone{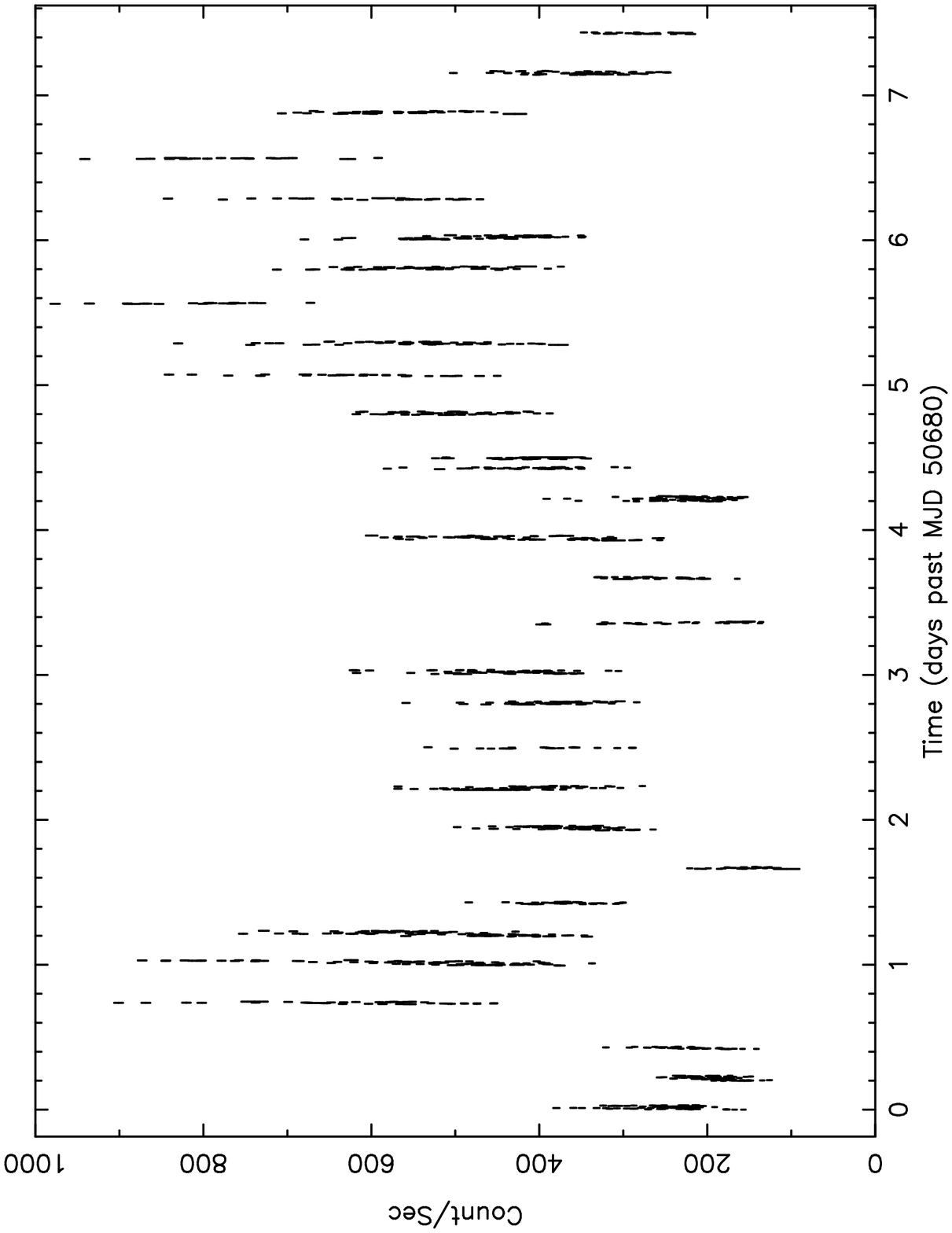}
\end{figure}
\newpage
\clearpage
\begin{figure}
\plotone{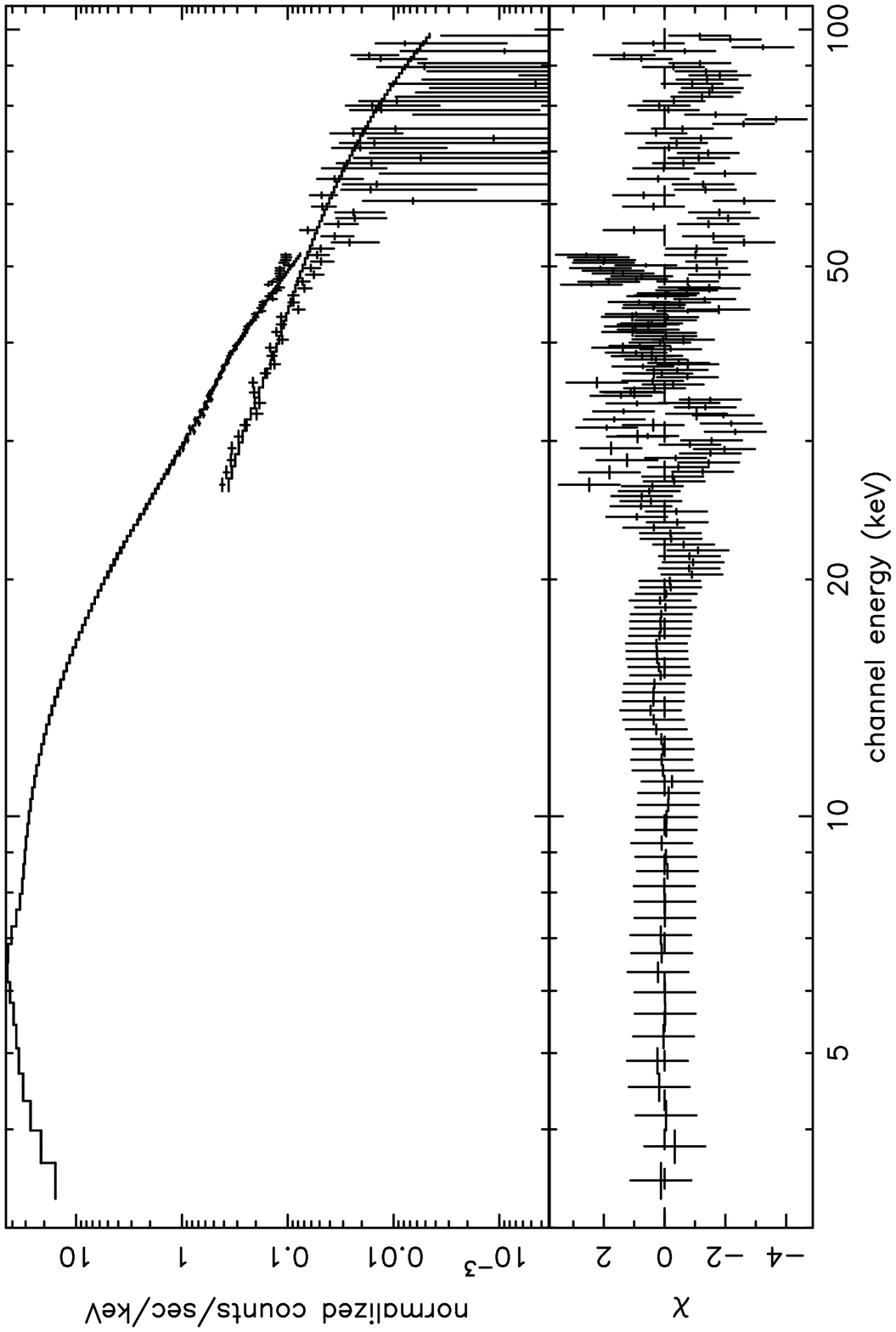}
\end{figure}
\newpage
\clearpage
\begin{figure}
\plotone{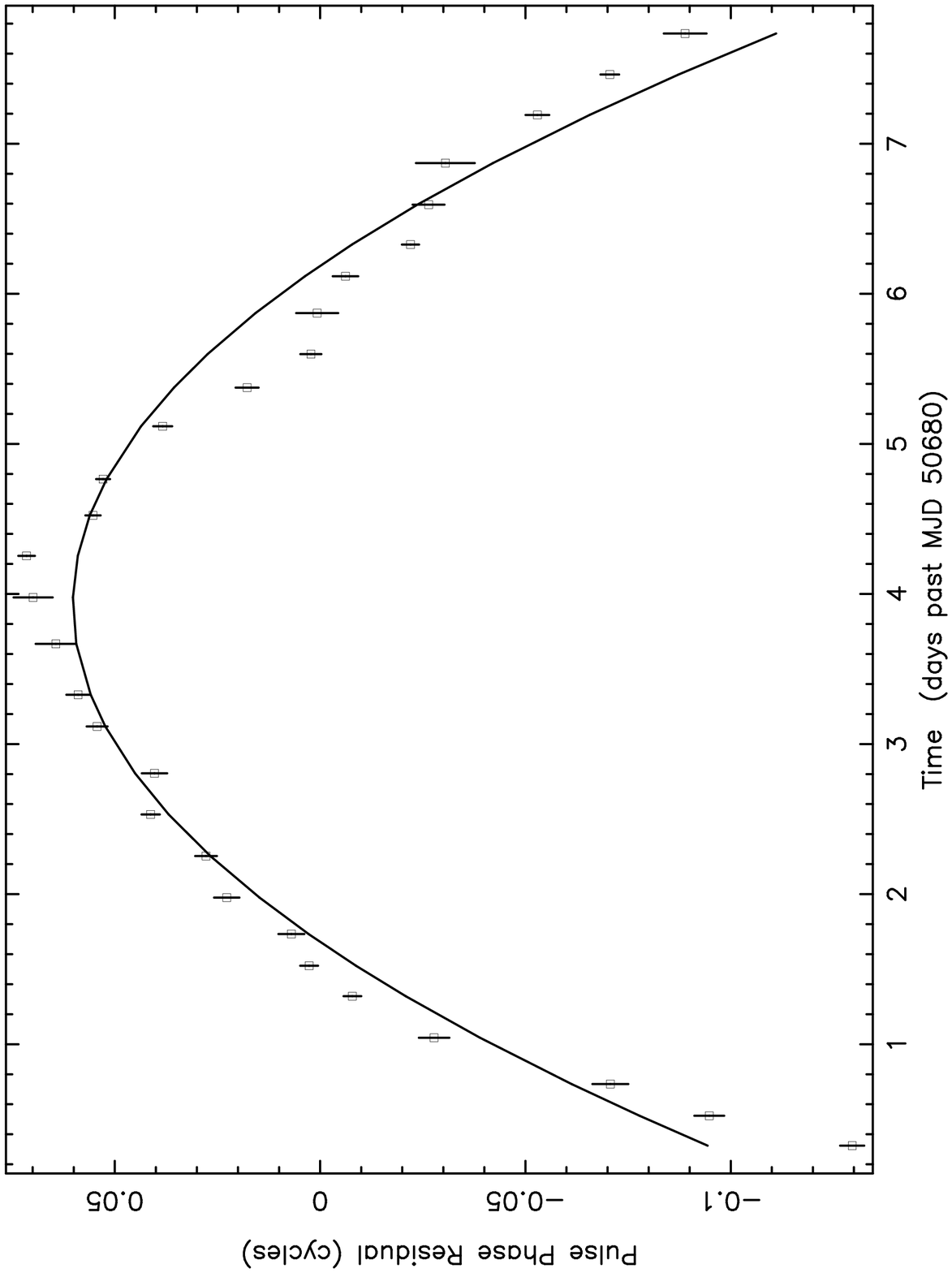}
\end{figure}
\newpage
\clearpage
\begin{figure}
\plotone{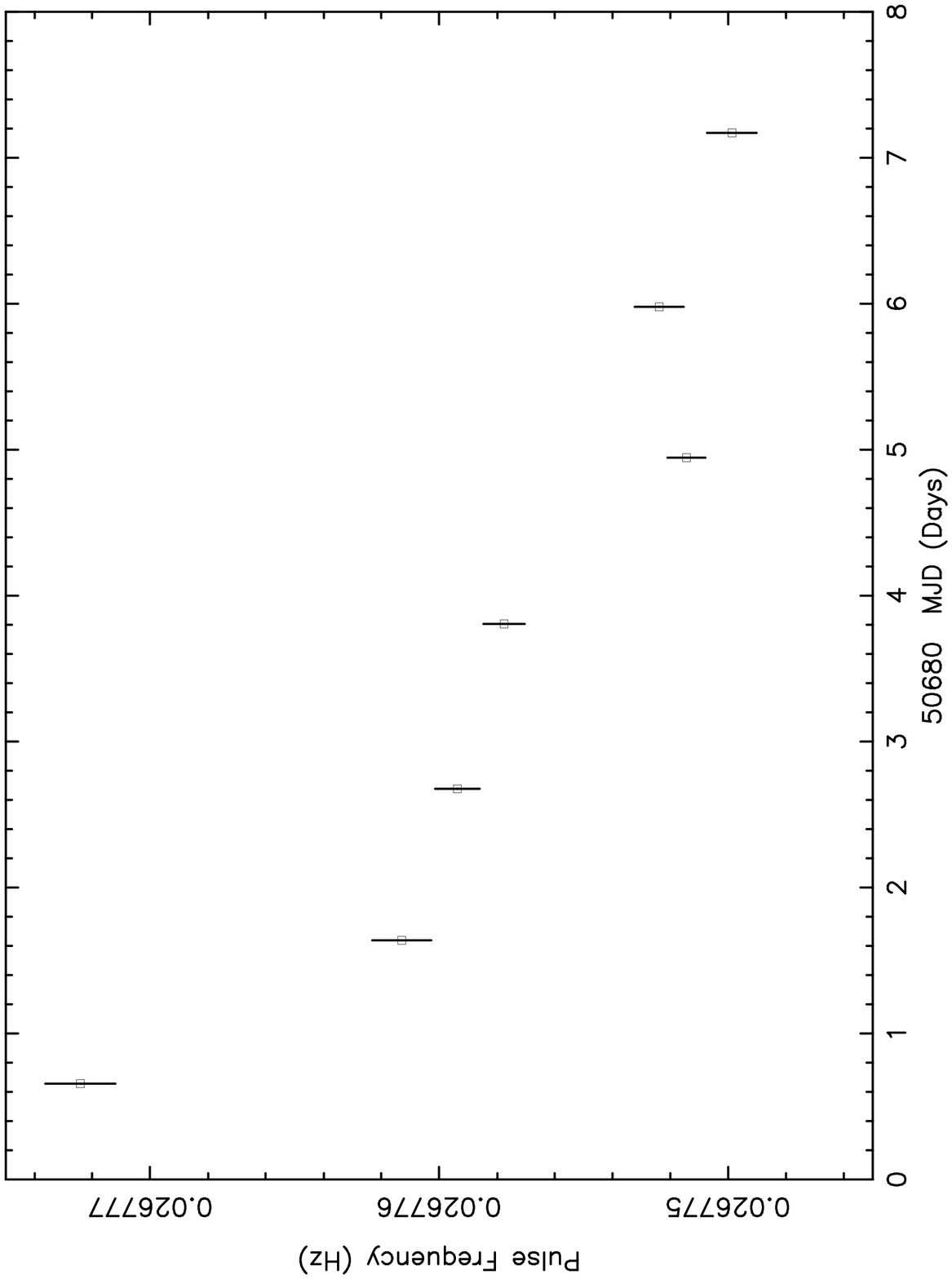}
\end{figure}
\newpage
\clearpage
\begin{figure}
\plotone{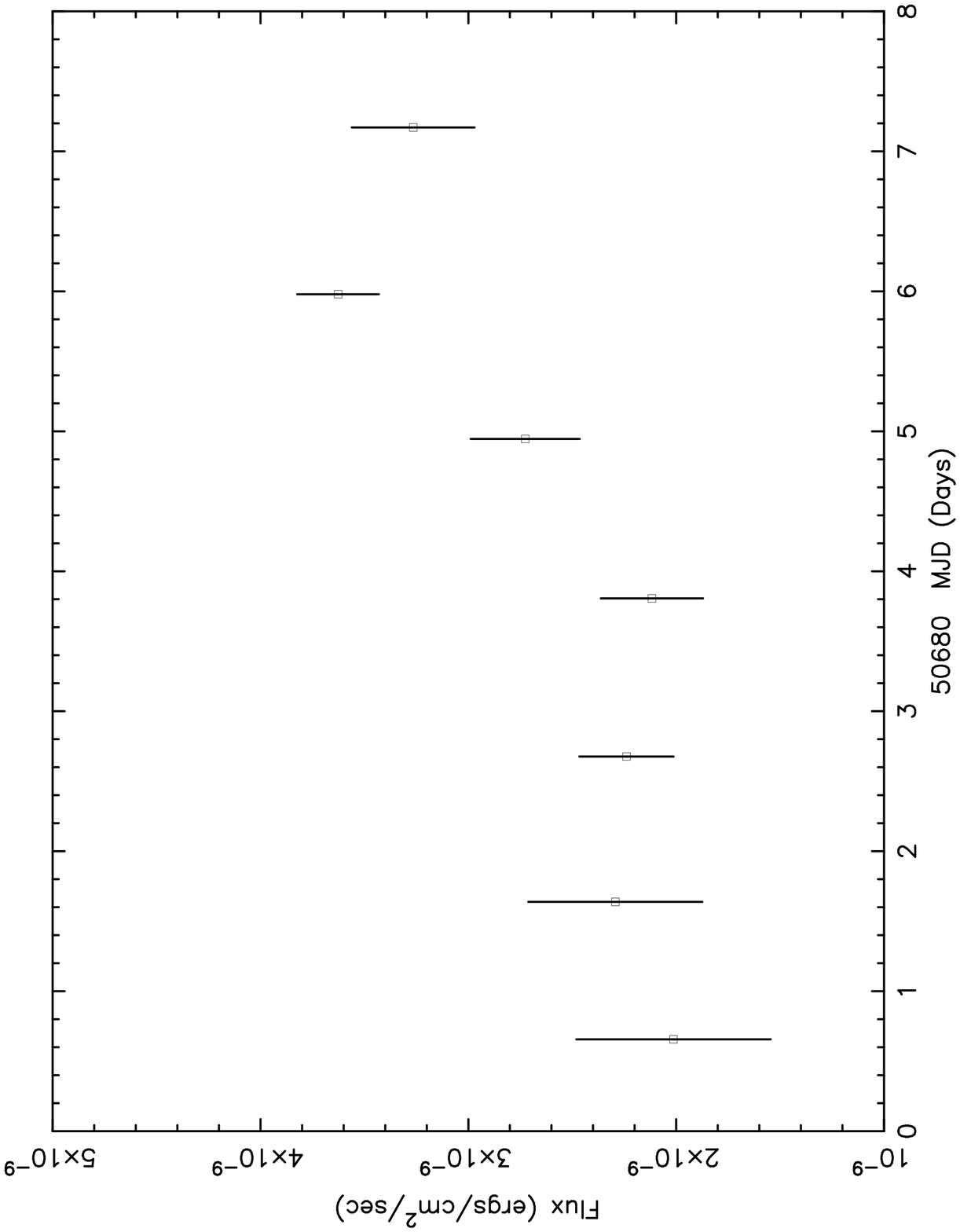}
\end{figure}


\begin{thebibliography}{}




\bibitem{} Baykal, A., 1997, A\&A, 319, 515


\bibitem{} Byrne, P., et al., 1981, ApJ 246, 951 

\bibitem{} Bildsten, L., et al.,
           1997, ApJS., 113, 367 

\bibitem{} Chakrabarty, D.,
           et al., 1993,  ApJ 403, L33

\bibitem{} Chakrabarty, D., et al.,
           1997, ApJ., 481, 101

                443, 786

\bibitem{} Deeter, J.E., Boynton, P.E.,Pravdo, S.H.,
           ApJ.,1981, 247, 1003

\bibitem{} Deeter, J.E., Boynton, P.E., 1985,
                in Proc. Inuyama Workshop on Timing
                Studies of X-Ray Sources, ed. S. Hayakawa
                $\&$ F. Nagase (Nagoya: Nagoya Univ.), 29 

\bibitem{} Ghosh, P., Lamb, F.K., 1979, ApJ., 232, 259 



\bibitem{} Gilfanov, M., Sunyaev, R., Churazov, E., Babalyan, G.,
           Pavlinskii, M., Yamburenko, N., Khavenson, N., 1991,
           Soviet Astron. Lett., 17, 46

\bibitem{} Jahoda, K., Swank, J., Giles, A.B.,
                Stark, M.J., Strohmayer, T., Zhang, W.,
                1996, Proc. SPIE, 2808, 59 



\bibitem{}  Kamata, Y., Koyama, K., Tawara, Y., Makishima, K.,
            Ohashi, T., Kawai, N., Hatsukade, I., 1990, PASJ
            42, 785







\bibitem{} Mereghetti, S., et al., 1991, ApJ 366, L23 

\bibitem{} Nelson, R.W., Bildsten, L., Chakrabarty, D., et al.,
                1997, 488, L117


\bibitem{}  Polidan, R.S., Pollard, G.S.G., Sanford, P.W., Locke, M.C.,
                1978, Nature 275, 296

\bibitem{} Rothschild, R.E., et al., 1998, 
           ApJ., 496, 538 

\bibitem{} Sunyaev, R., Gilfanov, M., Goldurm, A.,
           Schmitz-Frayesse, M.C.,
           1991, IAU Circ., No. 5342

\bibitem{} White, N.E., Pravdo, S.H., 1979, ApJ 233, L121



\end{thebibliography}
\end{document}